\documentclass[twocolumn,pdflatex,sn-nature,referee]{sn-jnl}

\usepackage{geometry}
\usepackage{graphicx}%
\usepackage{multirow}%
\usepackage{amsmath,amssymb,amsfonts}%
\usepackage{amsthm}%
\usepackage{mathrsfs}%
\usepackage[title]{appendix}%
\usepackage{xcolor}%
\usepackage{colortbl}
\usepackage{textcomp}%
\usepackage{manyfoot}%
\usepackage{booktabs}%
\usepackage{algorithm}%
\usepackage{algorithmicx}%
\usepackage{algpseudocode}%
\usepackage{listings}%

\theoremstyle{thmstyleone}%
\theoremstyle{thmstyletwo}%

\theoremstyle{thmstylethree}%

\begin{document}

\title[Article Title]{Multilingual Brain Decoding Reveals Shared Cortical Semantic Geometry}


\author[1,2]{\fnm{Yi} \sur{Guo}}
\author[3]{\fnm{Yong} \sur{Hu}}
\author[4]{\fnm{Michael Kwok-Po} \sur{Ng}}
\author*[1,2]{\fnm{Shuqiang} \sur{Wang}}

\affil[1]{\orgdiv{Shenzhen Institutes of Advanced Technology}, \orgname{Chinese Academy of Sciences}, \orgaddress{\city{Shenzhen}, \country{China}}}

\affil[2]{\orgname{University of Chinese Academy of Sciences}, \orgaddress{\city{Beijing}, \country{China}}}

\affil[3]{\orgname{The University of Hong Kong}, \orgaddress{\city{Hong Kong}, \country{China}}}

\affil[4]{\orgname{Hong Kong Baptist University}, \orgaddress{\city{Hong Kong}, \country{China}}}

\abstract{Human languages differ markedly in lexical and syntactic form yet can convey similar semantic content. The neural basis of this cross-language semantic commonality remains unclear. Here we developed a multilingual brain decoding framework and analysed brain recordings from 227 participants spanning Chinese, English and French. We found that cortical semantic representations exhibited consistent semantic geometry relationships across languages and individuals, revealing a shared cortical semantic geometry. Importantly, the cross-language consistency of cortical semantic geometry increased from sensorimotor to higher-order association cortex, forming a shared semantic geometry gradient closely aligned with the principal cortical functional gradient. Our work establishes generalizable multilingual brain decoding and identifies stable cortical semantic geometry as a neural basis for semantic commonality across languages.}

\maketitle

\section*{Introduction}\label{sec1}

\begin{figure*}[htbp]
    \centering
    \includegraphics[width=\textwidth]{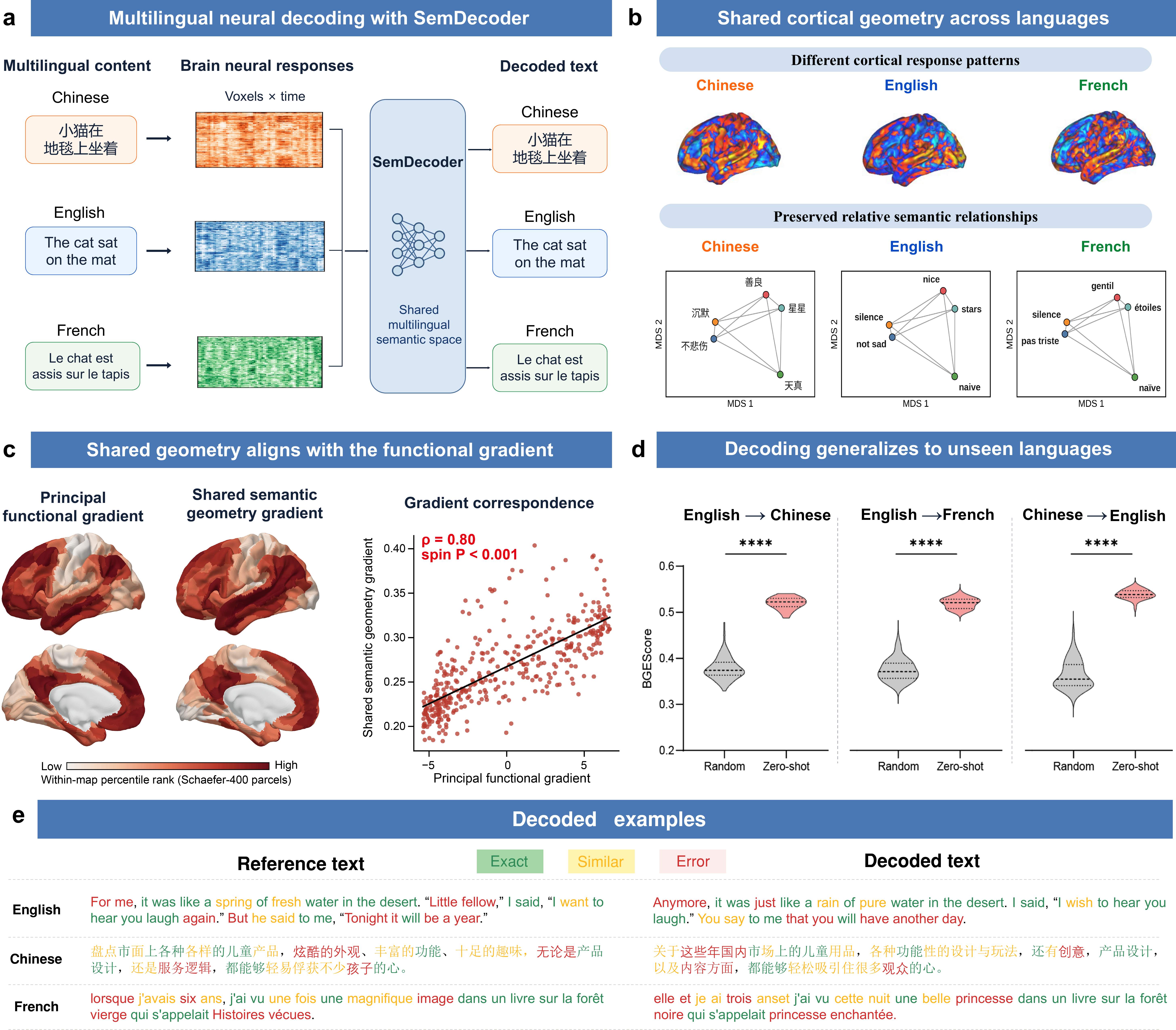}
    \caption{\textbf{Overview of multilingual brain decoding and shared cortical semantic geometry.} \textbf{a}, SemDecoder maps neural responses to Chinese, English and French stimuli into a unified semantic space for text reconstruction. \textbf{b}, Illustration of different cortical response patterns and preserved relative semantic relationships across languages. Corresponding semantic contents are shown in two-dimensional representational configurations. \textbf{c}, Cortical maps of the principal functional gradient and shared semantic geometry gradient, with their correspondence across Schaefer-400 parcels. Surface colours indicate percentile ranks within each map. Each point in the scatter plot denotes a parcel; the black line shows a least-squares fit. The Spearman correlation and one-sided spin-test $P$ value from 1,000 rotations are indicated. \textbf{d}, Zero-shot cross-language text decoding for English to Chinese, English to French and Chinese to English. BGEScore compares SemDecoder decoding with a control in which the model generates text without brain data (Random). Violin plots show participant distributions; dashed and dotted lines indicate medians and quartiles, respectively. \textbf{e}, Reference texts and corresponding decoded examples in English, Chinese and French. Green, yellow and red indicate exact matches, semantically similar content and unmatched content, respectively. ****$P<0.0001$.}
    \label{fig:1}
\end{figure*}

Human communication spans a remarkable diversity of languages that differ widely in lexical and syntactic form \cite{brochhagen2026meaning}. Across this diversity, different linguistic systems can convey comparable semantic content \cite{thompson2020cultural,fu2026semantic,lewis2023local}. Understanding how the brain preserves semantic commonality across languages is central to explaining the relationship between language and human cognition \cite{tuckute2024driving,jamali2024semantic,RN42,chen2017unified}.

Previous work has identified common features of cortical language networks across diverse languages \cite{RN6,RN66} and cross-language correspondences in neural representations of semantic content \cite{chen2026bilingual,yan2026shared}. Yet these findings do not establish whether the relationships among cortical semantic representations are preserved across languages and individuals. Representational geometry can capture these relationships through the similarities and dissimilarities among cortical neural representations of different semantic contents \cite{chen2017unified}. Determining whether this cortical geometry is preserved across languages and individuals despite differences in linguistic form and neural activity is therefore central to explaining the neural basis of semantic commonality.

To investigate the neural basis of cross-language semantic commonality, we proposed SemDecoder, a multilingual brain decoding framework that maps neural activity across languages and individuals into a unified semantic space for semantic decoding (Fig. \ref{fig:1}). Using functional magnetic resonance imaging (fMRI) recordings from 227 participants spanning Chinese, English and French, we examined whether semantic relationships are preserved in cortical response patterns across languages and individuals and how their consistency relates to cross-language decoding generalization.

We reveal a shared cortical semantic geometry across languages and individuals, characterized by preserved geometric relationships among cortical semantic representations. Crucially, the cross-language consistency of this geometry increased from sensorimotor regions towards higher-order association cortex, forming a shared semantic geometry gradient closely aligned with the principal cortical functional gradient. Along this gradient, the proposed framework showed stronger generalization to languages unseen during training, while cross-language neural response prediction was more accurate in regions with greater geometric consistency. Cross-language transfer also improved the decoding performance of the proposed framework with limited target-language data, while cortical regions with greater semantic geometry consistency contributed more strongly to transfer decoding. Overall, our work identifies shared cortical semantic geometry as a neural representational basis for semantic commonality across languages.

\section*{Results}\label{sec2}

\subsection*{Semantic decoding across languages and datasets}\label{subsec:generalizable-decoding}

\begin{figure*}[htbp]
    \centering
    \includegraphics[width=\textwidth]{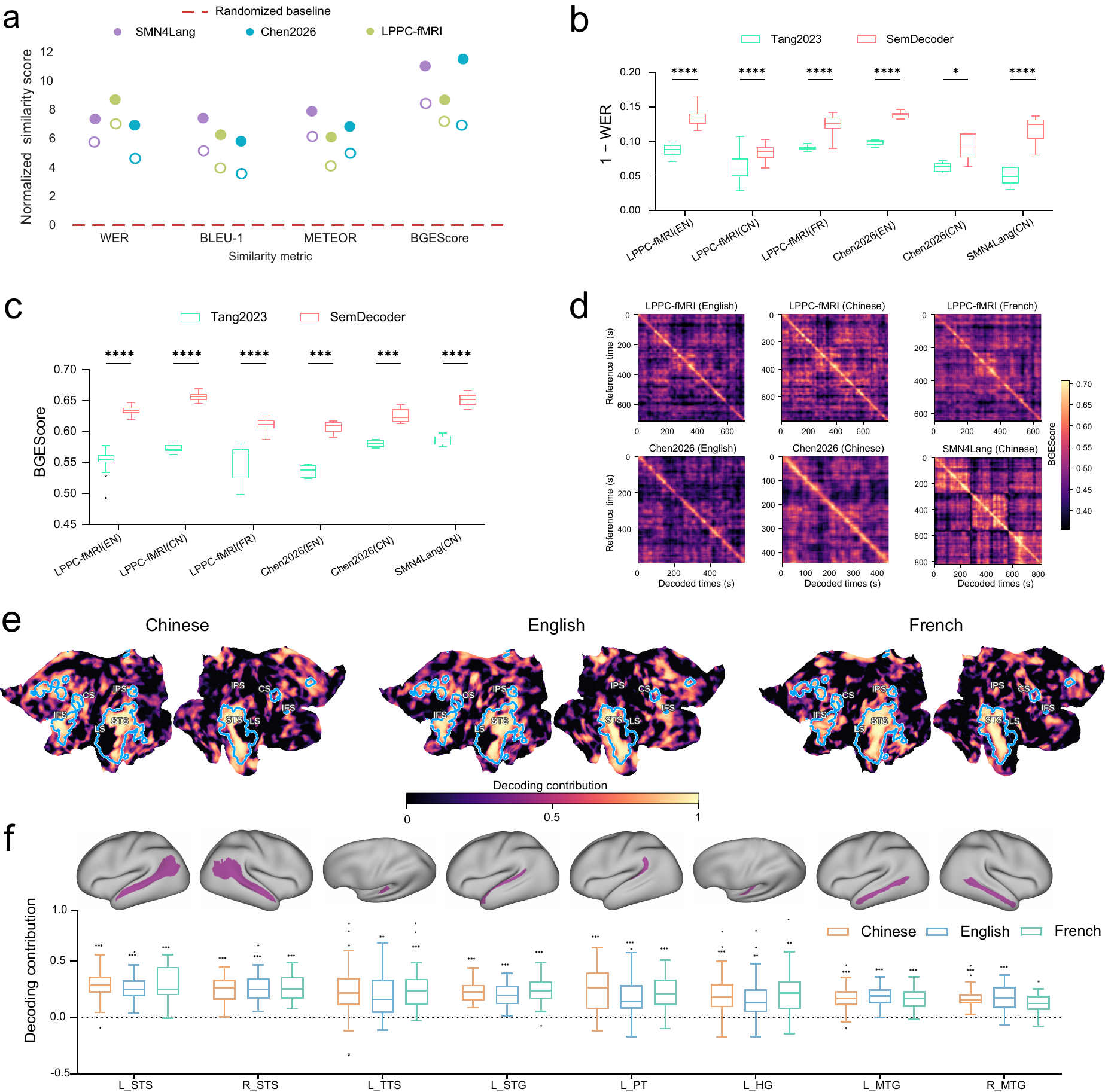}
    \caption{\textbf{Semantic decoding across languages and datasets.} \textbf{a}, Decoding performance of SemDecoder and Tang2023 \cite{RN7} across datasets. Open and filled circles indicate Tang2023 and SemDecoder, respectively; colours indicate datasets. Scores are $z$-normalized relative to text generated without brain data and oriented so that higher values indicate better performance; dashed line, randomized baseline. \textbf{b,c}, Participant-level $1-\mathrm{WER}$ (\textbf{b}) and BGEScore (\textbf{c}) across datasets and languages. $n=49$, 35, 28, 6, 6 and 12 participants for LPPC-fMRI(EN), LPPC-fMRI(CN), LPPC-fMRI(FR), Chen2026(EN), Chen2026(CN) and SMN4Lang(CN), respectively. Comparisons in \textbf{b,c} used two-sided paired $t$-tests with Holm--\v{S}id\'ak correction across the six comparisons for each metric; asterisks indicate adjusted $P$ values. \textbf{d}, Temporal BGEScore matrices comparing decoded texts with reference texts, averaged across participants. Horizontal and vertical axes indicate decoded and reference times, respectively. \textbf{e}, Cortical contribution to within-language semantic decoding, quantified by the decrease in brain--semantic cosine similarity after 8-mm searchlight occlusion. Colours show percentile ranks among positively contributing vertices; non-positive contributions are black. White lines mark major sulci. \textbf{f}, Participant-level contributions in eight temporal cortical regions, measured as the decrease in decoding performance after regional occlusion; insets identify each region. Asterisks indicate higher contributions than participants' cortex-wide means (one-sided paired Wilcoxon signed-rank tests with Holm correction). Dotted line, zero contribution. $n=35$, 49 and 28 participants for Chinese, English and French, respectively. Boxes in \textbf{b,c,f} show medians and interquartile ranges (IQRs); whiskers extend to $1.5\times$IQR, and points beyond whiskers denote outliers. *$P<0.05$, **$P<0.01$, ***$P<0.001$, ****$P<0.0001$. L, left; R, right; STS, superior temporal sulcus; TTS, transverse temporal sulcus; STG, superior temporal gyrus; PT, planum temporale; HG, Heschl's gyrus; MTG, middle temporal gyrus. EN, English; CN, Chinese; FR, French.}
    \label{fig:r1}
\end{figure*}

We first examined SemDecoder's ability to decode semantic content from brain activity across languages within a unified semantic space. We evaluated the accuracy and temporal correspondence of decoded content in Chinese, English and French (Fig. \ref{fig:r1}).

Whole-cortex decoding recovered semantically relevant content across all datasets and languages (Fig. \ref{fig:r1}). Representative decoded texts illustrated the recovery of words and semantic content from the references (Fig. \ref{fig:1}e). Performance exceeded the randomized baseline across the four evaluation metrics, and SemDecoder outperformed Tang2023 on every metric and dataset (Fig. \ref{fig:r1}a). At the participant level, SemDecoder achieved a lower word error rate (WER) and higher BGEScore than Tang2023 in each dataset and language (Fig. \ref{fig:r1}b,c). Temporal similarity matrices showed that decoded content corresponded to reference content at matching time points (Fig. \ref{fig:r1}d), indicating that decoding preserved the temporal alignment of semantic content.

Cortical contribution maps showed similar distributions across Chinese, English and French, with prominent contributions in bilateral temporal cortex (Fig. \ref{fig:r1}e). Regional summaries identified consistent contributions in superior and middle temporal regions across languages (Fig. \ref{fig:r1}f). Together, these findings show that SemDecoder decoded semantic content with temporal correspondence across the tested languages and datasets, with similar contributions from bilateral temporal cortex.

\subsection*{Cortical semantic geometry is shared across languages}\label{subsec:shared-geometry}

\begin{figure*}[htbp]
    \centering
    \includegraphics[width=\textwidth]{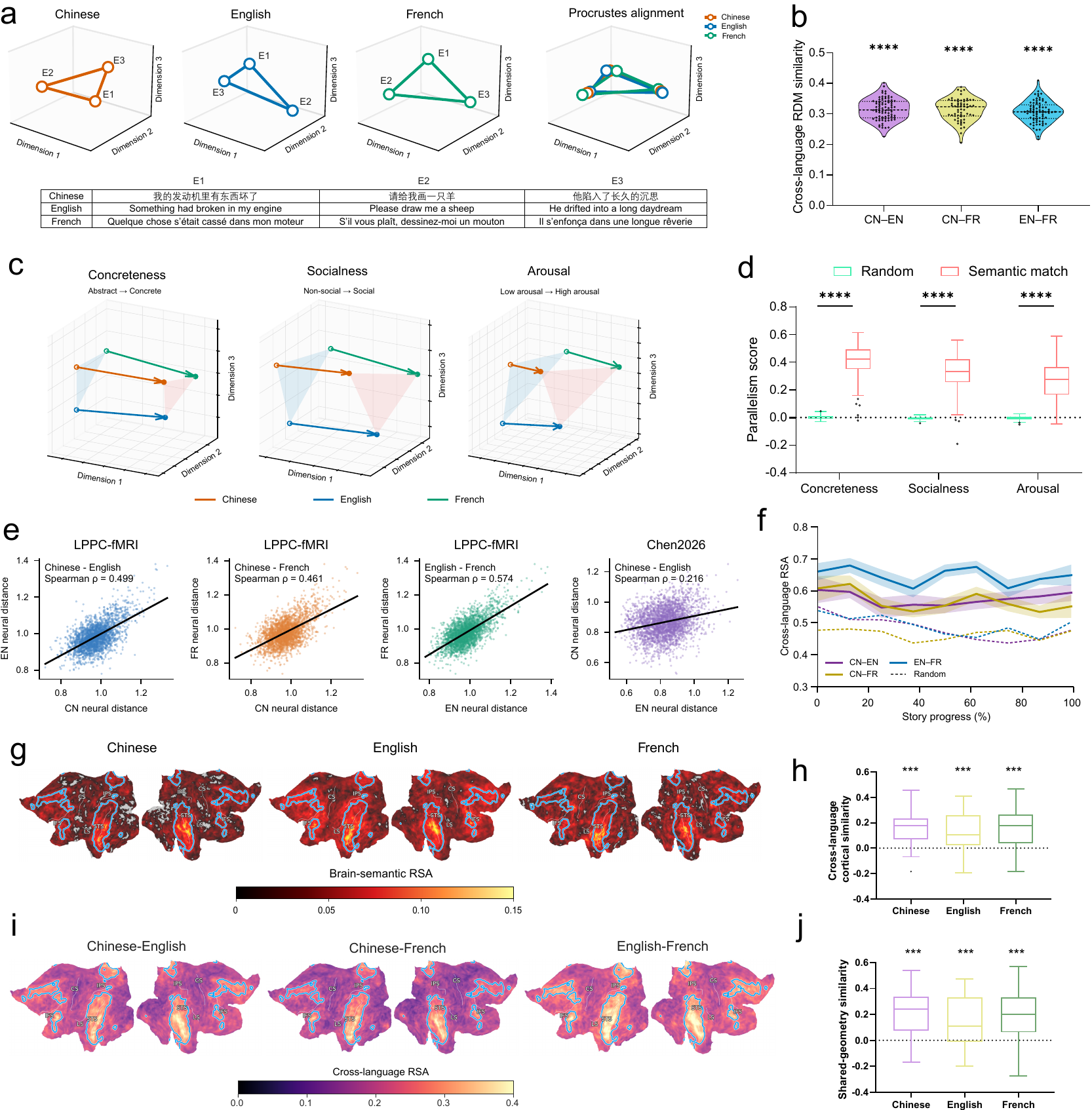}
    \caption{\textbf{Cortical semantic geometry is shared across languages.} \textbf{a}, Group-level cortical representational geometry before and after Procrustes alignment. Points denote corresponding narrative events (E1--E3), with distances reflecting dissimilarity between cortical response patterns. Similar relative configurations illustrate preserved relationships among events across languages. \textbf{b}, Cross-language RDM similarity between participants and the other language's group average, pooled across both directions for each language pair. Points denote participants; dashed and dotted lines indicate medians and quartiles, respectively. \textbf{c}, Concreteness, socialness and arousal directions in the aligned cortical representational spaces. Arrows point from abstract to concrete, non-social to social and low to high arousal, respectively. Parallel arrows across languages indicate consistent semantic attribute directions. \textbf{d}, Cross-language parallelism of attribute directions versus participant-specific median random controls ($n=112$ participants; one-sided paired Wilcoxon signed-rank tests, Holm-corrected across three attributes). \textbf{e}, Correspondence of neural distances between matched event pairs across languages. Points denote event pairs; lines show least-squares fits. Spearman correlations are indicated. \textbf{f}, Cross-language RSA along narrative progress. Solid lines and shading show the mean and s.e.m. across windows grouped by narrative progress; dashed lines show the phase-randomized control. \textbf{g}, Searchlight brain--semantic RSA for each language. \textbf{h}, Spatial Spearman correlations between individual brain--semantic RSA maps and the group averages of the other two languages, averaged across those languages. \textbf{i}, Searchlight cross-language RSA for each language pair. Cyan contours in \textbf{g,i} outline the top 20\% of mean cross-language RSA. \textbf{j}, Spatial Spearman correlations between individual brain--semantic and mean cross-language RSA maps. In \textbf{h,j}, $n=35$, 49 and 28 participants for Chinese, English and French, respectively. Asterisks in \textbf{h,j} indicate correlations above participant-specific spin-null medians (one-sided Wilcoxon signed-rank tests, Holm-corrected across languages within each panel). Boxes in \textbf{d,h,j} show medians and interquartile ranges (IQRs); whiskers extend to $1.5\times$IQR, and points beyond whiskers denote outliers. ***$P<0.001$, ****$P<0.0001$. s.e.m., standard error of the mean; CN, Chinese; EN, English; FR, French.}
    \label{fig:r2}
\end{figure*}

The recovery of semantic content across languages raised the question of whether semantic relationships were preserved in cortical response patterns across languages and individuals (Fig. \ref{fig:r2}). We compared cortical representational geometry across Chinese, English and French and related it to semantic content. The LPPC-fMRI language groups comprised different participants, allowing correspondence to be assessed across both languages and individuals.

Group-level neural representational dissimilarity matrices (RDMs) were computed separately for the three languages and projected into a low-dimensional space by multidimensional scaling. After Procrustes alignment, the relative configurations of semantic contents largely coincided across Chinese, English and French (Fig. \ref{fig:r2}a). Each participant's neural RDM also showed positive similarity to the group RDM of another language across all language pairs (Fig. \ref{fig:r2}b), extending the correspondence in cortical semantic geometry to individual participants.

Directions associated with concreteness, socialness and arousal were similarly oriented across languages (Fig. \ref{fig:r2}c). Their cross-language parallelism exceeded that obtained with random assignments of sentences to groups (Fig. \ref{fig:r2}d). Neural distances between matched event pairs were positively correlated across languages in LPPC-fMRI and between Chinese and English in Chen2026 (Fig. \ref{fig:r2}e). Semantic contents that were farther apart in one language also tended to be farther apart in another. Cross-language representational similarity analysis (RSA) correlations in sliding windows remained above the phase-randomized control throughout narrative progress (Fig. \ref{fig:r2}f), showing that shared representational structure persisted as the narrative unfolded.

Searchlight RSA comparing cortical and semantic dissimilarity matrices showed similar topographies across the three languages (Fig. \ref{fig:r2}g). Individual brain--semantic RSA maps also corresponded spatially to the group maps of the other languages (Fig. \ref{fig:r2}h). Direct cross-language searchlight RSA revealed similar spatial distributions of shared semantic geometry across language pairs, overlapping with regions showing stronger brain--semantic RSA (Fig. \ref{fig:r2}i). Individual brain--semantic maps also corresponded spatially to the mean shared semantic geometry map (Fig. \ref{fig:r2}j).

Across languages and individuals, cortical responses preserved both the relative distances among semantic contents and the directions associated with semantic attributes. Shared cortical semantic geometry thus captured a common organization of meaning across different linguistic forms.

\subsection*{Cross-language generalization follows a shared semantic geometry gradient}\label{subsec:unseen-languages}

\begin{figure*}[htbp]
    \centering
    \includegraphics[width=\textwidth]{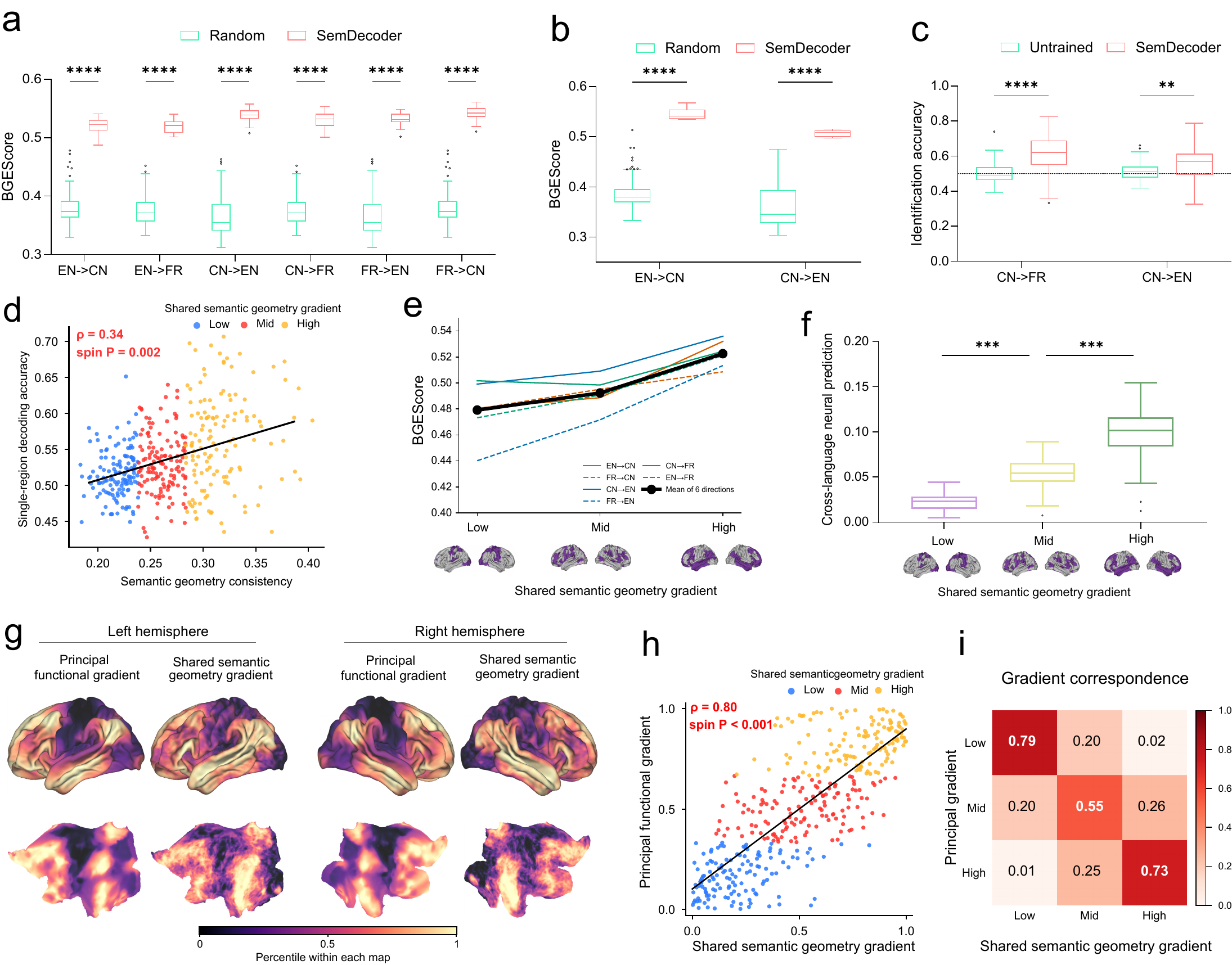}
    \caption{\textbf{Cross-language generalization follows a gradient of shared cortical semantic geometry.} \textbf{a}, Cross-language text decoding in LPPC-fMRI across all six language directions. BGEScore compares SemDecoder with text generated without brain data (Random). Arrows indicate training to test language. \textbf{b}, Bidirectional Chinese--English decoding in Chen2026. \textbf{c}, Decoders trained on LPPC-fMRI Chinese data were tested in NEBULA101 native French speakers listening to French ($n=97$) or English ($n=79$). An untrained network served as the control. Passage identification was measured by normalized rank among 12 candidates; dotted line, chance (0.5). \textbf{d}, Semantic geometry consistency versus single-region cross-language identification accuracy, averaged across six directions. Each point denotes a Schaefer-400 parcel. Identification uses two-alternative forced-choice matching with one parcel retained at a time. \textbf{e}, Cross-language text decoding with cortical inputs restricted to the low, middle or high third of the shared semantic geometry gradient for each language pair. Insets illustrate the Chinese--English groups. \textbf{f}, Cross-language neural response prediction (Pearson $r$) in low, middle and high thirds of the shared semantic geometry gradient ($n=112$ source participants, each averaged over two target languages). Comparisons used two-sided paired Wilcoxon signed-rank tests with Holm correction across three pairwise comparisons. \textbf{g}, Principal functional gradient, extending from sensorimotor to transmodal cortex, and shared semantic geometry gradient. Colours indicate percentile ranks within each gradient. \textbf{h}, Correspondence between the gradients across 400 cortical parcels, coloured by semantic geometry consistency group. Lines in \textbf{d,h} show least-squares fits; Spearman correlations and $P$ values are indicated. \textbf{i}, Overlap between independently defined thirds of the two gradients. Cells give the fraction of parcels in each functional-gradient third assigned to a third of the shared semantic geometry gradient; rows sum to one. Boxes in \textbf{a--c,f} show medians and interquartile ranges (IQRs); whiskers extend to $1.5\times$IQR, and points beyond whiskers denote outliers. **$P<0.01$, ***$P<0.001$, ****$P<0.0001$. CN, Chinese; EN, English; FR, French.}
    \label{fig:r3}
\end{figure*}

Regional variation in shared semantic geometry raised the question of whether its cortical organization was associated with cross-language generalization (Fig. \ref{fig:r3}). We evaluated decoding in languages whose neural data were excluded from decoder training and examined how generalization varied along the shared semantic geometry gradient and how this gradient related to the principal functional gradient.

Decoders trained on one language and evaluated on another performed above the randomized baseline in all directional pairs among Chinese, English and French in LPPC-fMRI (Fig. \ref{fig:r3}a). Bidirectional Chinese--English decoding in Chen2026 showed the same advantage within bilingual participants (Fig. \ref{fig:r3}b). Generalization extended to the independent NEBULA101 cohort \cite{rampinini2025nebula101}: decoders trained on LPPC-fMRI Chinese data identified passages heard in French or English more accurately than an untrained network (Fig. \ref{fig:r3}c). Semantic information learned from source-language neural responses therefore remained accessible across both languages and cohorts.

Cross-language semantic geometry consistency varied continuously across cortex, defining a shared semantic geometry gradient. Single-region cross-language identification accuracy was positively associated with a parcel's position along this gradient (Fig. \ref{fig:r3}d; Extended Data Fig. \ref{fig:ex_lppc_geometry_decoding}). To examine this relationship at the level of decoded text, we divided cortical parcels into low, middle and high thirds of the corresponding language-pair gradient and retained inputs from one third at a time. Mean BGEScore across language directions increased from the low to the high third (Fig. \ref{fig:r3}e). Cross-language neural response prediction showed a corresponding pattern: participant-level prediction accuracy increased from the low to the high third of the shared semantic geometry gradient (Fig. \ref{fig:r3}f; Extended Data Fig. \ref{fig:ex2}). Thus, the gradient organized cross-language generalization in both semantic decoding and neural response prediction.

The cortical distribution of the shared semantic geometry gradient closely followed the principal functional gradient, with semantic geometry consistency increasing from sensorimotor systems towards transmodal association cortex (Fig. \ref{fig:r3}g). Their correspondence was evident across cortical parcels (Fig. \ref{fig:r3}h; Spearman $\rho=0.80$). Independently defined thirds of the two gradients showed the greatest overlap at corresponding levels, particularly at their low and high ends (Fig. \ref{fig:r3}i), linking the organization of shared semantic geometry to the cortical functional hierarchy. This correspondence was stronger than for the other functional gradients (Extended Data Fig. \ref{fig:ex1}) and persisted after reliability adjustment. The association between semantic geometry consistency and decoding exceeded the null obtained by permuting cross-language sentence matches and was also supported by comparisons within levels of the principal functional gradient (Extended Data Fig. \ref{fig:ex3_functional_gradient_controls}). Neurosynth decoding further distinguished the gradients' apices, associating the shared semantic geometry gradient with auditory, speech and language terms and the principal functional gradient with theory of mind and self-referential processing (Extended Data Fig. \ref{fig:ex_neurosynth}).

The shared semantic geometry gradient linked the preservation of cortical semantic relationships to cross-language generalization, with stronger decoding and neural response prediction towards its high end. Its alignment with the principal functional gradient situated this relationship within the cortical functional hierarchy.

\subsection*{Cross-language transfer improves decoding with limited target-language data}\label{subsec:neural-information-sharing}

\begin{figure*}[htbp]
    \centering
    \includegraphics[width=\textwidth]{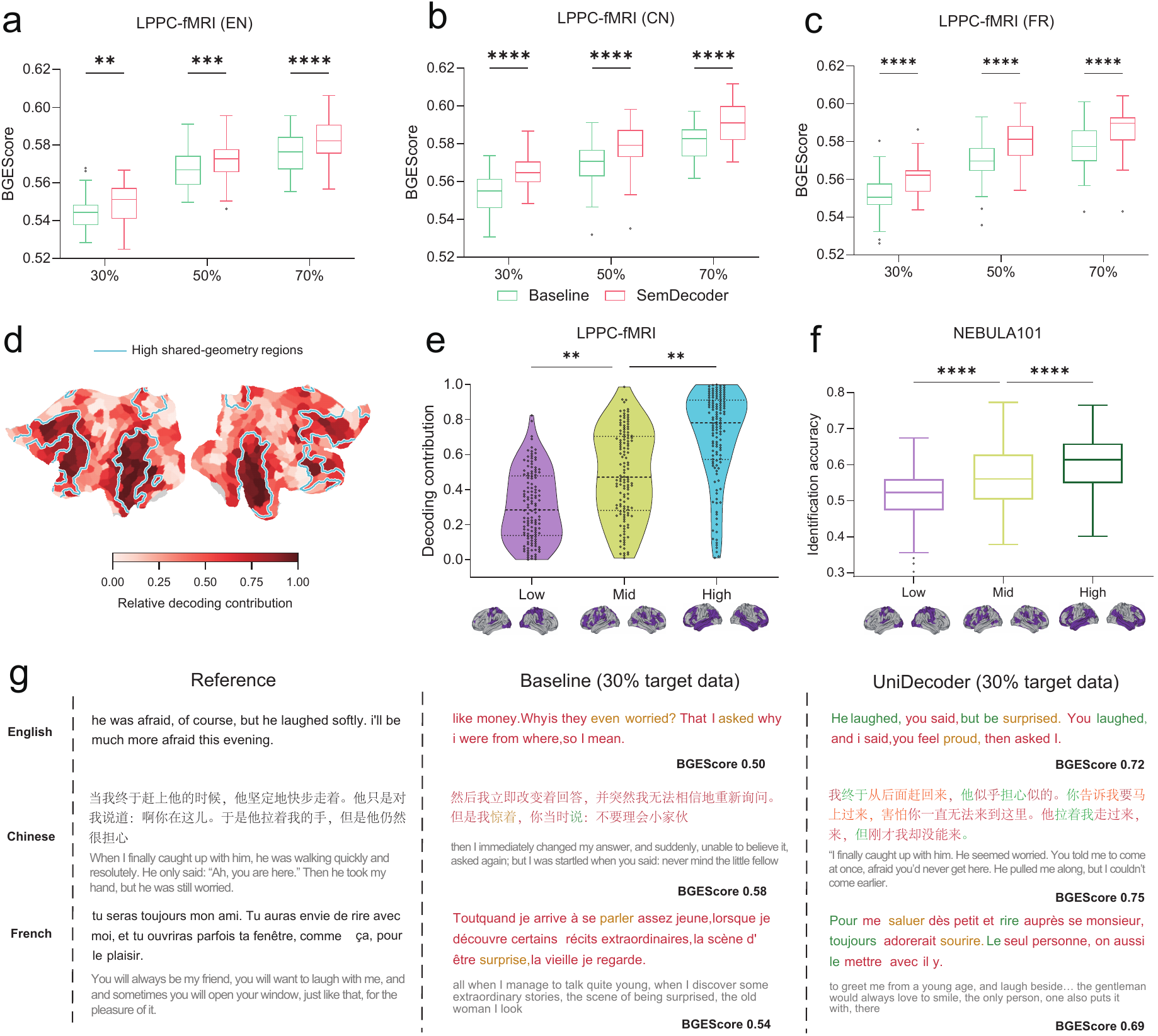}
    \caption{\textbf{Cross-language transfer improves decoding with limited target-language data.} \textbf{a--c}, Cross-subject cross-language transfer in LPPC-fMRI, with English (\textbf{a}), Chinese (\textbf{b}) and French (\textbf{c}) as the target language. Boxes compare BGEScore for cross-language transfer models and a baseline fitted to the same target-language data. Comparisons used two-way repeated-measures ANOVA with \v{S}id\'ak-adjusted comparisons between models at each proportion (three comparisons per target language). \textbf{d}, Relative cortical contribution to decoding by transfer models fitted with 30\% target-language data, measured by the reduction in brain--semantic cosine similarity after parcel occlusion. Colours show percentile ranks across Schaefer-400 parcels. Cyan contours mark the upper third of the shared semantic geometry gradient. \textbf{e}, Distributions of the relative contributions in \textbf{d} across low, middle and high thirds of the shared semantic geometry gradient. Points denote parcels ($n=133$, 133 and 134, respectively); dashed and dotted lines indicate medians and quartiles. Mid $>$ Low and High $>$ Mid were assessed using one-sided spin tests (1,000 rotations), with Holm correction across the two comparisons; asterisks indicate adjusted $P$ values. \textbf{f}, 12-way normalized-rank passage identification after cross-language transfer in native French speakers from NEBULA101 ($n=97$), using cortical inputs from low, middle or high semantic geometry consistency groups; chance, 0.5. Comparisons used two-sided paired Wilcoxon signed-rank tests with Holm correction across three pairwise comparisons. Insets in \textbf{e,f} illustrate the semantic geometry consistency groups. \textbf{g}, Reference passages and texts decoded by the baseline and SemDecoder using 30\% target-language training data. Green and orange highlight exact matches and semantically related content, respectively; red marks unmatched decoded content. Boxes in \textbf{a--c,f} show medians and interquartile ranges (IQRs); whiskers extend to $1.5\times$IQR, and points beyond whiskers denote outliers. **$P<0.01$, ***$P<0.001$, ****$P<0.0001$. EN, English; CN, Chinese; FR, French.}
    \label{fig:r4}
\end{figure*}

Having established generalization without target-language neural training data, we examined whether cross-language semantic transfer could improve decoding when target-language training data were limited. We compared cross-language transfer models with a baseline fitted to the same restricted target-language data, then examined cortical contributions and decoding along the shared semantic geometry gradient.

In LPPC-fMRI, SemDecoder achieved higher BGEScore than the target-language baseline for English, Chinese and French at every tested data proportion, including the smallest training set (Fig. \ref{fig:r4}a--c). Both models improved as more target-language data became available, while transfer retained an advantage throughout.

Cortical contributions to decoding after cross-language transfer increased from the low to the high third of the shared semantic geometry gradient (Fig. \ref{fig:r4}d,e). In the independent NEBULA101 cohort, passage identification after transfer likewise improved when decoding used cortical inputs from successively higher thirds of this gradient (Fig. \ref{fig:r4}f). Representative decoded passages illustrated the advantage over the target-language baseline: English, Chinese and French decoded texts retained more of the reference content after transfer than when decoding relied on limited target-language data alone (Fig. \ref{fig:r4}g).

Cross-language semantic transfer improved decoding with limited target-language data. The shared semantic geometry gradient was associated with decoding after transfer as well as zero-shot generalization, with its high end showing stronger cortical contributions and better passage identification.

\section*{Discussion}\label{sec3}

Our study reveals shared cortical semantic geometry across languages, showing that geometric relationships among semantic contents are preserved in cortical activity during language processing (Fig. \ref{fig:r2}). The cross-language consistency of cortical semantic geometry increases from sensorimotor to higher-order association cortex, forming a shared semantic geometry gradient closely aligned with the principal cortical functional gradient. Cross-language decoding generalization also increases along this gradient (Fig. \ref{fig:r3}). When target-language neural training data are limited, cross-language semantic transfer improves decoding performance, and cortical contributions to transfer decoding increase along the shared semantic geometry gradient (Fig. \ref{fig:r4}).

The human capacity to understand and communicate comparable meanings through different languages is fundamental to communication and provides an important perspective on the relationship between language and conceptual cognition \cite{lambonralph2017semantic,fu2026semantic}. At the same time, cross-linguistic studies have revealed systematic differences in word meanings associated with linguistic and cultural experience \cite{thompson2020cultural,lewis2023local}. Understanding how the human brain supports semantic commonality across languages despite these differences is therefore important for uncovering the neural basis of human language. Similarities in linguistic behaviour alone cannot establish whether relationships among semantic contents are also preserved in cortical activity. Previous studies have identified shared cortical language networks and common neural semantic representations across languages, while recent recordings from hippocampal neurons have revealed shared semantic geometry \cite{RN6,chen2026bilingual,yan2026shared}. Despite these advances, a key question remains unresolved: how shared semantic relationships are represented across the cerebral cortex and whether their cross-language consistency follows the cortical functional hierarchy. Our findings address this question by identifying preserved geometric relationships among semantic contents in cortical activity (Fig. \ref{fig:r2}) and a systematic gradient in their cross-language consistency (Fig. \ref{fig:r3}).

The shared cortical semantic geometry identified in our study highlights a level of cross-language neural commonality that extends beyond similarities in cortical activity patterns associated with individual semantic contents. Geometric relationships among semantic representations can remain comparable even when the underlying neural activity patterns differ, making representational geometry particularly informative for investigating shared semantic relationships across languages \cite{kriegeskorte2013representational,kriegeskorte2021neural}. Our analyses revealed cross-language correspondence in neural distances among semantic events, while representational directions associated with different semantic attributes showed significant cross-language parallelism. Cross-language representational similarity also persisted across successive time windows during ongoing language processing (Fig. \ref{fig:r2}). These converging findings indicate that cross-language semantic commonality is reflected not only in the representation of individual semantic contents but also in the geometric relationships among them. The preservation of these relationships provides a common representational structure through which semantic contents can retain comparable relationships across languages despite differences in cortical activity patterns.

The shared semantic geometry gradient further reveals a systematic cortical hierarchy in the cross-language preservation of semantic relationships. Previous studies have identified a principal cortical functional gradient extending from sensorimotor systems to transmodal association cortex, reflecting a major axis of macroscale cortical organization \cite{margulies2016gradient,smallwood2021default}. We found that cross-language semantic geometry consistency increases along a closely corresponding cortical axis (Fig. \ref{fig:r3}). This finding echoes previous studies describing hierarchical and timescale-dependent aspects of speech and language processing \cite{RN68,caucheteux2023evidence,yamashita2025conversational}. Whereas these studies characterized different levels of language processing and their cortical distributions, our findings reveal a hierarchy in the preservation of semantic relationships across languages. This pattern suggests that higher-order association cortex may support semantic relationships that are less dependent on specific linguistic forms. Neurosynth meta-analysis further showed that cortical parcels ranked in the highest 5\% of the shared semantic geometry gradient had stronger associations with auditory, speech and language terms than parcels independently selected from the highest 5\% of the principal functional gradient (Extended Data Fig. \ref{fig:ex_neurosynth}). This functional difference suggests that the shared semantic geometry gradient is not simply a reproduction of the principal cortical functional gradient but exhibits a particularly strong association with language-related processing.

Cross-language decoding provides further evidence for the functional relevance of shared cortical semantic geometry. The decoding framework generalized across languages without target-language neural recordings being used for supervised brain-to-semantic mapping, with decoding performance increasing towards the high-consistency end of the shared semantic geometry gradient (Fig. \ref{fig:r3}). Cross-language neural response prediction showed a corresponding association with semantic geometry consistency, consistent with previous work linking neural population geometry to abstraction and generalization \cite{bernardi2020geometry,wakhloo2026geometry}. When target-language neural training data were limited, transfer from other languages improved decoding performance relative to models trained using the same amount of target-language data alone. Cortical contributions to transfer decoding also increased along the shared semantic geometry gradient (Fig. \ref{fig:r4}). These improvements under limited-data conditions have broader implications for equity in multilingual brain decoding. Neural training data are not equally available across languages, and approaches that require extensive recordings for each target language may place speakers of less well-resourced languages at a disadvantage. Differences in data availability could therefore translate into disparities in decoding performance and access to emerging neurotechnologies. By allowing neural data from other languages to support decoding with limited target-language recordings, cross-language semantic transfer offers a potential means of reducing these barriers \cite{singh2025transfer,spalding2026shared,RN15}. Although fairness across linguistic populations was not directly evaluated in the present study, this capacity for data-efficient transfer may help mitigate disparities arising from unequal neural data resources and support the development of more equitable multilingual brain–computer interfaces \cite{RN75}.

The present study examined a limited range of languages, and the generality of shared cortical semantic geometry across human languages remains to be established. Extending this research to more typologically diverse languages could determine how broadly shared semantic relationships are preserved across linguistic diversity \cite{RN6,fu2026semantic}. Language experience represents another important direction for future research. Studies involving individuals with different multilingual experiences could reveal how language learning and use influence cross-language semantic geometry consistency, providing further insight into the plasticity of shared semantic representations \cite{mechelli2004plasticity,costa2014bilingual}. Ultimately, extending the study of shared cortical semantic geometry across linguistic diversity and language experience may deepen our understanding of how the human brain supports common semantic understanding across different languages, offering a broader perspective on the relationship between human language and cognition.

\section*{Methods}\label{sec4}
\subsection*{Problem formalization}
The decoding task is to reconstruct natural language text that captures the meaning of a stimulus presented to a participant. Each stimulus sentence is represented as a word sequence \( W = (w_1, \dots, w_L) \), which is tokenized by a pretrained multilingual model (PMM) into a sequence of \( N \) tokens \( \mathcal{S} = (s_1, \dots, s_N) \). The PMM computes semantic embeddings for all tokens, forming a matrix \( \mathbf{Y} \in \mathbb{R}^{N \times d} \), where \( d \) is the hidden-state dimension of the selected PMM layer.

The corresponding fMRI responses form a neural signal matrix \( \mathbf{X} \in \mathbb{R}^{t \times c} \), where \( t \) and \( c \) denote the number of time points and cortical grayordinates, respectively.

The brain recordings \( \mathbf{X} \) are mapped into the unified semantic space, yielding a predicted semantic matrix \( \hat{\mathbf{Y}} \in \mathbb{R}^{N \times d} \) aligned with \( \mathbf{Y} \). From \( \hat{\mathbf{Y}} \), a decoded token sequence \( \tilde{\mathcal{S}} = (\tilde{s}_1, \dots, \tilde{s}_N) \) is generated using a semantic consistency-guided beam search algorithm. This sequence is then detokenized by the PMM into the final decoded word sequence \( \tilde{W} = (\tilde{w}_1, \dots, \tilde{w}_{\tilde{L}}) \).

\subsection*{Datasets}
Four publicly available fMRI datasets from healthy participants were analysed. All datasets were collected under institutional ethics approval. Key characteristics, including participant numbers, languages and recording durations, are summarized in Extended Data Table \hyperref[tab:E1]{\textbf{1}}.

LPPC-fMRI contains fMRI recordings from adults presented with English, Chinese and French versions of \textit{The Little Prince} \cite{RN8}. Ethical approval for this research was granted by the ethics committees of Cornell University, Jiangsu Normal University, and the French regional biomedical research ethics committee.

In Chen2026, fluent Chinese--English bilingual participants read natural narratives in their native Chinese and non-native English during fMRI \cite{chen2026bilingual}. The study was approved by the Committee for Protection of Human Subjects at the University of California, Berkeley, and all participants provided informed consent.

SMN4Lang participants listened to Mandarin news broadcasts during fMRI \cite{RN10}. This study received approval from the Institutional Review Board of Peking University.

For NEBULA101, we used fMRI recordings from the French and English intact-passage conditions \cite{rampinini2025nebula101}. Participants whose native language was French listened to intact passages from \textit{Alice's Adventures in Wonderland} in French ($n=97$), with 79 of these participants also contributing data in the English condition. Each participant heard 12 passages of approximately 18 s in each language condition. Data collection was approved by the Geneva Cantonal Ethics Commission.

\subsection*{Data preprocessing}
The fMRI data were preprocessed using the ABCD--HCP protocol, implemented through the ABCD-BIDS pipeline \cite{RN40} and based on the Human Connectome Project (HCP) workflow \cite{RN29}. Preprocessing included denoising, motion and distortion correction, spatial registration, brain segmentation, cortical surface reconstruction and projection to fsLR-32k surface space. Whole-cortex decoding used all 59,412 cortical grayordinates in fsLR-32k surface space. To account for the haemodynamic delay, blood-oxygen-level-dependent (BOLD) responses were sampled for each stimulus token at seven delays from 0 to 12 s after token onset, in steps of 2 s. Signals at these time points were estimated by linear interpolation between adjacent fMRI acquisitions. The responses across delays were concatenated as decoder inputs and integrated using learned attention weights.

Aligned stimulus transcripts for the primary datasets were obtained from their original releases. Intact auditory passages from the French and English NEBULA101 conditions were transcribed using Whisper large-v3 \cite{RN45}. The stimulus language was specified during transcription, with temperature set to zero, a beam size of five and conditioning on previous text disabled. Semantic features were extracted with BLOOM-1.1B \cite{RN27}, a PMM supporting 46 languages. {Semantic features were derived from the hidden representations of the 20th layer of BLOOM-1.1B, using a 15-token context window per segment. These 20th-layer embeddings served as the target semantic representations for training SemDecoder, and the same PMM layer was used during semantic consistency-guided decoding.}

The three primary datasets were divided into mutually exclusive training and test sets at the story level. One or more complete stories were reserved for testing, while the remainder were used for training. The training and evaluation partitions are summarized in Extended Data Table \hyperref[tab:E2]{\textbf{2}}.

\subsection*{SemDecoder}
SemDecoder maps fMRI recordings from different subjects into a shared semantic space and reconstructs text from the resulting representations through semantic consistency-guided beam search.

\par\addvspace{0.5\baselineskip}
\noindent\textbf{Neural-to-semantic mapping.} Inter-individual variability in neural responses complicates multi-subject brain decoding \cite{RN7}. For subject \(u\) and stimulus-aligned sample \(i\), the cortical responses at \(L_{\mathrm{n}}\) haemodynamic delays, \(\mathbf{X}_{u,i}\in\mathbb{R}^{L_{\mathrm{n}}\times c}\), are first mapped into a shared latent space through a subject-specific linear projection followed by feature-wise scaling and shifting:
\begin{equation}
\widetilde{\mathbf{X}}_{u,i} = \left(\mathbf{X}_{u,i}\mathbf{W}_{u} + \mathbf{b}_{u}\right)\odot\left(\mathbf{1}+\boldsymbol{\gamma}_{u}\right) + \boldsymbol{\beta}_{u},
\label{eq:harmonization}
\end{equation}
\noindent where \(\mathbf{W}_{u}\in\mathbb{R}^{c\times p}\) and \(\mathbf{b}_{u},\boldsymbol{\gamma}_{u},\boldsymbol{\beta}_{u}\in\mathbb{R}^{p}\) are learnable parameters of subject \(u\). Each subject's projection thereby defines a participant-specific pathway into the shared space, and all subsequent operations are shared across subjects and languages. A shared neural feature extractor captures the spatiotemporal structure of \(\widetilde{\mathbf{X}}_{u,i}\), yielding \(\mathbf{H}_{u,i}\in\mathbb{R}^{L_{\mathrm{n}}\times p}\). The feature extractor is pretrained on designated pretraining datasets through masked neural modelling, in which a subset of neural tokens is withheld and predicted from the remaining spatiotemporal context without semantic supervision. Attention pooling integrates \(\mathbf{H}_{u,i}\) across delays, and a projection head maps the pooled representation into the \(d\)-dimensional space of the 20th hidden layer of BLOOM-1.1B. The resulting vectors form the predicted semantic matrix \(\hat{\mathbf{Y}}\in\mathbb{R}^{N\times d}\). We set \(L_{\mathrm{n}}=7\) and \(p=256\).

\par\addvspace{0.5\baselineskip}
\noindent\textbf{Training objective.} SemDecoder is trained to align predicted semantic representations with standardized PMM features of the stimulus text. The loss combines a symmetric contrastive loss \cite{infoNCE} and a cosine similarity loss with equal weights:
\begin{equation}
\mathcal{L} = \mathcal{L}_{\text{NCE}} + \mathcal{L}_{\text{CS}},
\label{eq:loss}
\end{equation}
\begin{equation}
\mathcal{L}_{\text{CS}} = \frac{1}{B}\sum_{i=1}^{B}\left(1-\cos(\mathbf{y}_i,\hat{\mathbf{y}}_i)\right),
\label{eq:loss_cs}
\end{equation}
\begin{equation}
\mathcal{L}_{\text{NCE}}
 = -\frac{1}{2B}\sum_{i=1}^{B}
   \log\frac{e^{2s_{ii}}}
   {\left(\sum_{j=1}^{B}e^{s_{ij}}\right)
    \left(\sum_{j=1}^{B}e^{s_{ji}}\right)},
\label{eq:loss_nce}
\end{equation}
\noindent where \(s_{ij}=\cos(\hat{\mathbf{y}}_i,\mathbf{y}_j)/\tau\), \(\mathbf{y}_i\) and \(\hat{\mathbf{y}}_i\) are the target and predicted semantic vectors of sample \(i\), \(B\) is the number of samples in a mini-batch, the other samples in the mini-batch serve as negatives, and \(\tau\) is a learnable temperature.

\par\addvspace{0.5\baselineskip}
\noindent\textbf{Semantic consistency-guided beam search.} Text is generated from \(\hat{\mathbf{Y}}\) with beam search \cite{RN31}, following the framework in which a language model proposes candidates and neural data select among them \cite{RN7}. At each step, BLOOM-1.1B proposes the \(M\) most probable next tokens for each retained sequence, conditioned on its preceding 15 tokens. Candidate sequences are scored by their accumulated cosine similarity to the predicted semantic representations:
\begin{equation}
\operatorname{score}(\tilde{\mathcal{S}})
  = \sum_{i=1}^{N}
    \frac{\hat{\mathbf{y}}_i \cdot \tilde{\mathbf{y}}_i}
         {\lVert \hat{\mathbf{y}}_i \rVert \, \lVert \tilde{\mathbf{y}}_i \rVert},
  \qquad \tilde{s}_i \in \mathcal{C}_i ,
\label{eq:beam_score}
\end{equation}
\begin{equation}
\mathcal{C}_i
  = \underset{s \in \mathcal{V}}{\operatorname{top\text{-}}M}\;
    P\!\left(s \mid \tilde{s}_1, \ldots, \tilde{s}_{i-1}\right),
\label{eq:beam_candidates}
\end{equation}
\noindent where \(\mathcal{V}\) is the PMM vocabulary and \(\tilde{\mathbf{y}}_i\) is the layer-20 representation of the candidate sequence at token \(i\), standardized using training-story statistics. The language model therefore defines the candidate set, and selection among candidates is determined by the decoded neural representations. The \(k\) highest-scoring sequences are retained at each step, with \(k=12\) and \(M=96\). The final decoded word sequence \(\tilde{W}\) is obtained by detokenizing \(\tilde{\mathcal{S}}\) using the tokenizer of the PMM.

\subsection*{Cortical representational analyses}

Sentence-level cortical response patterns were estimated using least-squares-separate (LSS) general linear models (GLMs) \cite{mumford2012lss}. BOLD time series were $z$-standardized at each vertex within each story. A separate GLM was fitted for each target sentence, with one regressor for that sentence and another for all remaining sentences combined, together with an intercept and discrete cosine drift terms with a cutoff period of 128 s. Sentence events were modelled as boxcars spanning their actual onset and offset times and convolved with the SPM canonical haemodynamic response function. The target-sentence $\beta$ estimates formed the cortical response patterns used for subsequent representational analyses. The analysis included all 59,412 cortical grayordinates in fsLR-32k surface space.

Twenty-four evenly spaced sentences were selected from each story as anchor events. For the separate Chinese, English and French participant groups, corresponding sentence indices were used across versions of each story; for the bilingual reading data, anchors were matched according to normalized story progress. For each participant and story, a neural representational dissimilarity matrix (RDM) was constructed from the anchor-event response patterns using correlation distance. Language-level RDMs were obtained by averaging participant RDMs within each language, and cross-language correspondence was quantified using Spearman correlation between the vectorized upper triangles of corresponding RDMs. Group-level RDMs were embedded into three dimensions using classical multidimensional scaling and aligned using Procrustes analysis for visualization.

For individual cross-language correspondence, whole-cortex RDMs based on 24 anchors per story across the nine matched stories were compared with the other language's group RDM using Spearman correlation. Participant values from both directions were pooled for each language pair (Fig. \ref{fig:r2}b).

Brain--semantic and cross-language cortical correspondence were assessed using searchlight representational similarity analysis \cite{kriegeskorte2008rsa}. Neural RDMs were computed within 8-mm geodesic searchlights centred on each cortical grayordinate. Semantic RDMs were constructed from the corresponding anchor sentences using cosine distance between PMM layer-20 sentence representations. Brain--semantic correspondence was quantified using Spearman correlation between neural and semantic RDMs for each participant, story and searchlight, followed by Fisher transformation and averaging across stories. Cross-language cortical similarity was quantified by directly correlating group-level neural RDMs between language pairs at each searchlight. Statistical significance was assessed using permutation procedures described below.

\par\addvspace{0.5\baselineskip}
\noindent\textbf{Semantic attribute directions.} Semantic attribute scores were derived from published English norms for concreteness \cite{brysbaert2014concreteness}, socialness \cite{diveica2023socialness} and arousal \cite{warriner2013affective}. English sentences were lowercased and tokenized, and function words were removed using a predefined list. Content words were matched to the norms, with unmatched forms rechecked after stripping common suffixes or contracted endings. Concreteness ratings used only single-word entries. Sentence scores were the mean ratings of matched words; sentences without matches were excluded from grouping for that attribute. The same scores were assigned to the aligned Chinese and French sentences. Scores were adjusted for word count and log duration within stories. Whole-cortex sentence response patterns were standardized within stories and adjusted for story, word count and log duration. Sentences were assigned to lower and upper score thirds within each story. Attribute directions were defined as the upper-minus-lower centroid difference, pooled across stories. Participant-level parallelism was the mean cosine similarity between each participant's direction and the group directions of the other two languages. Random controls reassigned sentences to groups within stories over 200 draws.

\par\addvspace{0.5\baselineskip}
\noindent\textbf{Shared semantic geometry maps.} Searchlight RSA correlations were Fisher-transformed, averaged across stories and transformed back to correlation values for each language pair. The shared semantic geometry gradient was defined as
\begin{equation}
g_{\mathrm{sem}}(v)=\frac{\mathcal{R}_{\mathrm{CN,EN}}(v)+\mathcal{R}_{\mathrm{CN,FR}}(v)+\mathcal{R}_{\mathrm{EN,FR}}(v)}{3},
\label{eq:shared_geometry}
\end{equation}
\noindent where \(v\) denotes a cortical location and \(\mathcal{R}_{a,b}(v)\) is the story-averaged RSA correlation between languages \(a\) and \(b\). CN, EN and FR denote Chinese, English and French, respectively. The values of \(g_{\mathrm{sem}}\) express consistency on the correlation scale; cortical displays use within-map percentile ranks.

Cortical distribution maps used all nine matched stories. Functional-gradient comparisons and participant-level neural response prediction groups used the mean of the three language-pair maps estimated from the decoding training set (Extended Data Table \hyperref[tab:E2]{\textbf{2}}). Groups for text decoding used the map estimated from training data for the corresponding language pair, whereas transfer contribution groups used the mean Chinese--English and English--French training-set maps. NEBULA101 transfer groups used the three-pair mean map from all nine source stories, independently of the NEBULA101 data. Individual cortical correspondence was quantified by spatial Spearman correlation across Schaefer-400 parcels \cite{schaefer2018parcellation}. Each participant's brain--semantic RSA map was compared with the group maps of the other two languages, averaging the two correlations, and with the mean shared semantic geometry map.

\par\addvspace{0.5\baselineskip}
\noindent\textbf{Semantic geometry stability.} Stability across stories was assessed using 100 random splits of the nine stories into sets of four and five. For each language pair, an RSA map was computed separately for each set using Fisher averaging. Stability was quantified by the spatial Spearman correlation between the two maps. Stability along narrative time was assessed using group-level sentence RDMs constructed from all sentences matched across the three languages. Cross-language RSA was calculated within 30-sentence windows advanced by 10 sentences and aligned by normalized narrative progress. Circular-shift null distributions used 1,000 shifts of sentence order. The minimum circular displacement was the larger of 10 sentences and 10\% of story length, rounded up to an integer. A complementary null applied the same sentence-response estimation procedure to phase-randomized BOLD data. Each run used a common phase randomization across cortical locations. Windows were pooled across stories within 10\% bins of normalized narrative progress. Curves show the mean and s.e.m. across windows within each bin; bins containing fewer than five windows were omitted.

\par\addvspace{0.5\baselineskip}
\noindent\textbf{Cortical functional gradients.} The first three functional gradients (G1--G3) were obtained from the published reference maps of Margulies et al. \cite{margulies2016gradient} through neuromaps \cite{markello2022neuromaps}. G1 denotes the principal functional gradient, extending from sensorimotor to transmodal association cortex. Reference maps and fMRI data were represented in the same fsLR-32k cortical surface space. G1 was oriented towards higher values in default-mode regions; G2 and G3 retained their original signs. Gradient and semantic geometry values were averaged within Schaefer-400 parcels, excluding unassigned vertices. Correspondence was quantified by Spearman correlation between raw parcel values. Cortical displays used percentile ranks within each map. Functional-hierarchy groups were defined by ranking cortical locations or parcels along G1, as appropriate to each analysis.

\par\addvspace{0.5\baselineskip}
\noindent\textbf{Neurosynth meta-analytic decoding.} Association $z$-maps for 24 cognitive terms selected following Margulies et al. \cite{margulies2016gradient} were derived from Neurosynth v7 \cite{yarkoni2011neurosynth}, projected onto fsLR-32k and averaged within Schaefer-400 parcels. For each gradient, parcels were ranked into 20 equal-sized bins. Heatmaps show mean association $z$-statistics, with negative bin means set to zero; terms were ordered by their weighted mean position along the principal functional gradient. Apex enrichment was defined as the difference in mean association $z$-statistics between the top 5\% of parcels of the shared semantic geometry gradient and those of the principal functional gradient. To account for gradient correspondence, the rank-transformed semantic geometry map was regressed on the rank-transformed principal functional gradient at the vertex level. Residual maps were rotated 1,000 times using bilaterally coupled spherical rotations and added back to the fitted component before parcel averaging. Enrichment was recomputed for each surrogate, and one-sided $P$ values were calculated as $(k+1)/1{,}001$, where $k$ counts surrogate values at least as large as the observed enrichment.

\subsection*{Cross-language generalization and transfer}

\par\addvspace{0.5\baselineskip}
\noindent\textbf{Cross-language decoding.} Decoders trained on a source language were evaluated on brain responses to a target language whose neural data were excluded from decoder training. Evaluation used the held-out test sets specified in Extended Data Table \hyperref[tab:E2]{\textbf{2}}. Cross-language decoding covered all six directions among Chinese, English and French. Each target participant's standardized cortical responses were passed through every source-participant pathway, and the predicted semantic vectors were averaged across pathways before text generation. The bilingual reading analyses evaluated both Chinese--English directions using each participant's corresponding pathway. Decoded text was assessed using the language similarity metrics described below.

Regional analyses across the three language groups used standardized responses averaged across target-language participants and passed through the source-participant pathways. One Schaefer-400 parcel was retained at a time, with standardized inputs outside that parcel set to zero across temporal delays. Parcels containing fewer than 10 cortical vertices were excluded. Semantic identification was evaluated by two-alternative forced-choice matching of predicted and reference segment representations. A pair was correctly identified when the sum of its two matched cosine similarities exceeded the sum for the swapped pairing.

To assess decoding across levels of semantic geometry consistency, cortical parcels were divided into three approximately equal-sized groups using the map estimated from the corresponding language groups' training data. The whole-cortex source-language decoder was then evaluated on target-language group-mean responses from the held-out test set, retaining inputs from one cortical group at a time without retraining. Inputs outside the selected group, including unassigned vertices, were set to zero after standardization. Predictions were averaged across source-participant pathways before text generation.

\par\addvspace{0.5\baselineskip}
\noindent\textbf{Independent validation in bilingual participants.} The whole-cortex decoder trained on LPPC-fMRI Chinese recordings was applied to NEBULA101 French and English responses. Standardized cortical responses were passed through the source-participant pathways, and predicted semantic vectors were averaged across pathways and within each passage. Reference passage vectors were obtained from the corresponding transcripts using the same semantic feature extraction procedure. Passage identification used cosine similarity between predicted and reference vectors across the 12 passages heard by each participant. Normalized rank accuracy was the mean proportion of the other 11 candidates ranked below the correct passage, with chance performance of 0.5. Control scores were obtained from untrained networks with the same architecture and averaged over three random initializations.

For transfer analyses with restricted cortical inputs in NEBULA101, models initialized from the Chinese source-language decoder were fitted for ten epochs using six passages from 30\% of listeners. Models were tested on the other six passages of all listeners, with passage halves exchanged across two folds. Inputs outside each semantic geometry consistency group were set to zero during both fitting and testing. Cortical parcels were divided into three approximately equal-sized consistency groups using the three-pair mean semantic geometry map estimated from all nine source stories. Each listener's scores were combined across folds using the same 12-way normalized-rank metric for passage identification.

\par\addvspace{0.5\baselineskip}
\noindent\textbf{Cross-language neural response prediction.} Vertex-wise ridge models were fitted separately for each source participant in the three language groups to sentence-level LSS response estimates, without additional temporal delays. BLOOM layer-20 token representations were averaged within each sentence, retaining preceding within-run context. Features were reduced to 64 dimensions using principal component analysis (PCA); PCA and feature standardization were fitted on source-language training stories and applied to test features. Evaluation used nine-fold leave-one-story-out cross-validation. The regularization parameter was selected by nested leave-one-story-out cross-validation on source-participant responses averaged within 148 Destrieux regions. Cross-language predictions were compared with target-language group-mean responses; within-language predictions used the group mean excluding the source participant. Vertex-wise Pearson correlations across held-out sentences were Fisher-transformed and averaged across folds. For Fig. \ref{fig:r3}f, the two target-language maps were averaged in Fisher space within each source participant. Values were then averaged across all vertices in each semantic geometry consistency group and transformed back to correlations.

Neural response prediction models for Chen2026 were fitted to sentence-level responses and evaluated within each bilingual participant using leave-one-story-out cross-validation across all ten stories. Within-language and cross-language predictions were compared with that participant's held-out responses. Feature extraction, PCA dimensionality and regularization selection otherwise followed the same procedure.

\par\addvspace{0.5\baselineskip}
\noindent\textbf{Semantic geometry and generalization.} Regional associations between semantic geometry consistency and decoding performance were quantified using Spearman correlation across Schaefer-400 parcels. Cortical groups were defined from the training-set geometry maps described above. Spatial associations were evaluated using spin permutations \cite{alexanderbloch2018spin}.

Regional reliability was estimated within each language by correlating sentence RDMs from two participant halves. Reliability-adjusted associations residualized ranked semantic geometry consistency and G1 against the three language-specific reliability ranks and an intercept. Residual correlations were evaluated by rotating G1 while retaining reliability covariates. For the correspondence control, 1,000 permutations of cross-language sentence matches were generated within each training story. Three-pair mean semantic geometry maps were related to single-region identification accuracy averaged across six language directions (Extended Data Fig. \ref{fig:ex3_functional_gradient_controls}a,b).

\par\addvspace{0.5\baselineskip}
\noindent\textbf{Transfer with limited target data.} The conditions labelled 30\%, 50\% and 70\% used the first two, four and six target-language training stories in LPPC-fMRI, and the first three, four and six in Chen2026, respectively. Both baseline and transfer models started from the SMN4Lang-pretrained shared trunk. Transfer models additionally learned from source-language neural data before fitting the same target-language subset as the baseline. Source data came from participants in the other two language groups, or from the same bilingual participants' other-language recordings. Subject-specific projection, scaling and shifting parameters were newly initialized for the target participants. Separate target-language validation and test data were held out from fitting; both models were evaluated on the same target participants and test data.

\subsection*{Statistical analysis}

Comparisons between decoding methods in Fig. \ref{fig:r1}b,c used two-sided paired $t$-tests on one summary score per participant and method, with Holm--\v{S}id\'ak correction separately across the six comparisons for each metric.

Regional cortical contributions were compared with each participant's cortex-wide mean contribution using one-sided paired Wilcoxon signed-rank tests. The directional alternative tested whether regional contributions exceeded the cortex-wide mean. Holm correction was applied jointly across 148 Destrieux regions and three languages, comprising 444 comparisons.

Monte Carlo permutation-based $P$ values were calculated as $(k+1)/(N+1)$, where $k$ counts null statistics at least as extreme as the observed statistic and $N$ is the number of permutations. Cross-language correspondence between whole-cortex RDMs was assessed using 1,000 anchor-label permutations within stories. Participant-level brain--semantic RSA maps were assessed using 5,000 one-sided sign-flip permutations, with Benjamini--Hochberg false-discovery-rate correction across cortical locations within each language. Group-level brain--semantic and cross-language searchlight RSA maps were assessed using 1,000 and 300 anchor-label permutations, respectively. Labels were shuffled within each story, with the same permutation applied across cortical locations. Maximum-statistic null distributions provided family-wise error correction across the cortical map for each language or language pair.

Individual cross-language RDM similarities in Fig. \ref{fig:r2}b were tested against zero using one-sided Wilcoxon signed-rank tests, pooling participant values across the two directions within each language pair.

Semantic attribute parallelism (Fig. \ref{fig:r2}d) was compared with each participant's median score from 200 within-story random group assignments using one-sided paired Wilcoxon signed-rank tests ($n=112$), with Holm correction across the three attributes. For Fig. \ref{fig:r2}h,j, each participant's observed spatial correlation was compared with the median of their spin null distribution by rotating the reference group maps. These differences were tested using one-sided Wilcoxon signed-rank tests within each language, with Holm correction across the three languages separately for each panel.

Neural response prediction across gradient thirds (Fig. \ref{fig:r3}f) was compared using two-sided paired Wilcoxon signed-rank tests across 112 source participants, with Holm correction across all three pairwise comparisons. Transfer decoding scores (Fig. \ref{fig:r4}a--c) were analysed using two-way repeated-measures ANOVA with model and target-data proportion as factors. Baseline and transfer models were compared within each proportion using \v{S}id\'ak correction across the three comparisons separately for each target language. Passage identification across gradient thirds (Fig. \ref{fig:r4}f) was compared using two-sided paired Wilcoxon signed-rank tests across 97 listeners, with Holm correction across all three pairwise comparisons.

Neural response prediction maps were tested using one-sided sign flips of source participants' Fisher-transformed correlations. For the three language groups, 10,000 random sign flips and vertex-level max-$T$ controlled whole-cortex family-wise error at $P<0.05$. For the six bilingual participants, all 64 sign configurations were enumerated, with a cluster-forming threshold of $t>2.015$ and cluster-mass family-wise error correction at $P<0.05$. Spatial associations were evaluated using 1,000 spin permutations on the fsLR spherical surface. Regional geometry--decoding associations in Extended Data Fig. \ref{fig:ex_lppc_geometry_decoding} used one-sided tests across all 400 parcels, with Holm correction across the six language directions. For the three-gradient comparison (Extended Data Fig. \ref{fig:ex1}), the semantic geometry map was rotated and re-averaged within parcels; two-sided tests compared absolute correlations. The comparison with the principal functional gradient used a one-sided test. Reliability-adjusted associations rotated G1 while holding the reliability covariates fixed.

Group comparisons in Fig. \ref{fig:r4}e and Extended Data Fig. \ref{fig:ex3_functional_gradient_controls}c used one-sided spin tests with 1,000 rotations. Semantic geometry maps were rotated at the vertex level with bilaterally coupled rotations and averaged within Schaefer-400 parcels; parcels lacking valid rotated values were assigned the map mean. For Fig. \ref{fig:r4}e, each rotated map was divided at its one-third and two-thirds quantiles, and differences in mean contribution percentile were calculated for Mid minus Low and High minus Mid. For Extended Data Fig. \ref{fig:ex3_functional_gradient_controls}c, the original principal functional gradient thirds remained fixed. Within each third, parcels were regrouped above versus at or below the median rotated semantic geometry value, and the difference in mean identification accuracy was calculated as Higher minus Lower. Parcel contributions and identification accuracies remained fixed across rotations. Observed differences were compared with their respective null distributions using the Monte Carlo formula above. Holm correction was applied separately across the two comparisons in Fig. \ref{fig:r4}e and the three comparisons in Extended Data Fig. \ref{fig:ex3_functional_gradient_controls}c.

\subsection*{Cortical contribution analysis}
Cortical contributions to within-language semantic decoding were quantified by searchlight occlusion of the trained whole-cortex decoders for the Chinese, English and French language groups. The models were evaluated on the held-out test set. For each searchlight, its cortical input features were set to zero after standardization across all temporal delays, while inputs outside the searchlight were retained. Zero therefore represented the mean of the standardized features. Model parameters remained fixed throughout the analysis.

Contribution was defined as the decrease in mean token-level cosine similarity between predicted semantic vectors and standardized BLOOM layer-20 targets after occlusion. For group maps, predictions were averaged across participant-specific pathways before cosine similarity was calculated. Participant-level maps compared each participant's predictions with the targets before and after occlusion.

For visualization, positive searchlight contributions were converted to percentile ranks within each language's group map; non-positive contributions were displayed in black. Regional contributions were calculated separately for each participant as the decrease in the same decoding score after occluding each of the 148 Destrieux regions \cite{RN43}. These contributions were compared with each participant's cortex-wide mean contribution using the tests described in Statistical analysis.

Contributions to decoding after cross-language transfer (Fig. \ref{fig:r4}d,e) were estimated using the transfer models for the three language groups under the condition labelled 30\%, fitted to the target-language training subset after pretraining on the other two languages. Each Schaefer-400 parcel was occluded in turn by zeroing standardized inputs across all delays in the held-out test set. Predictions were averaged across target participants before calculating cosine similarity to standardized BLOOM layer-20 targets. Contribution was the decrease from the unoccluded score. Decreases were $z$-scored across parcels within language, averaged across the three languages and converted to percentile ranks as $(\mathrm{rank}-0.5)/400$. Groups and contours used the mean Chinese--English and English--French semantic geometry maps estimated from the training set.

\subsection*{Language similarity metrics}
Decoded texts were evaluated against reference texts using WER, BLEU-1 \cite{RN53} and METEOR \cite{RN54}. Sentence-level semantic similarity was measured using BGEScore, defined as the cosine similarity between multilingual BGE-M3 embeddings \cite{RN26} of decoded and reference texts. Comparisons with Tang2023 \cite{RN7} used identical data partitions and evaluation procedures.

The random baseline (Random) used the same semantic consistency-guided beam search as SemDecoder, with the cosine similarity score of each candidate replaced by a value drawn uniformly at random. All other settings, including the candidate set, beam width and sequence length, were identical. We generated 200 random sequences with different random seeds for each test text and scored each against the reference text using the same evaluation metrics.

Cross-metric comparisons used the normalized similarity metric $\mathcal{Z}_{\mathrm{sim}}$. Scores were \textit{z}-standardized relative to the empirical null distribution, with the null mean~$\mu$ and standard deviation~$\sigma$ estimated from the corresponding distribution of similarity scores:
\begin{equation}
	\mathcal{Z}_{\mathrm{sim}} = \frac{\mathrm{sim}(W,\hat{W}) - \mu}{\sigma},
	\label{eq:7}
\end{equation}
\noindent
where $\mathrm{sim}(\cdot,\cdot)$ denotes BLEU-1, METEOR or BGEScore. For WER, the direction was reversed: $\mathcal{Z}_{\mathrm{WER}}=(\mu_{\mathrm{WER}}-\mathrm{WER})/\sigma_{\mathrm{WER}}$. Thus, higher standardized values indicate better performance for all metrics.


\bibliography{sn-bibliography}

%
%
%

\setcounter{table}{0}
\renewcommand{\theHtable}{extended.\arabic{table}}
\renewcommand{\tablename}{}
\renewcommand{\thetable}{Extended Data Table \arabic{table}}

\setcounter{figure}{0}
\renewcommand{\theHfigure}{extended.\arabic{figure}}
\renewcommand{\figurename}{Extended Data Fig.}  
\renewcommand{\thefigure}{\arabic{figure}}  

\begin{table*}[t!]
\centering
\caption{Datasets.}
\renewcommand{\arraystretch}{1.2}
\setlength{\tabcolsep}{5.5pt}

\begin{tabular}{@{}llllll@{}}
\toprule
\textbf{Dataset}        & \textbf{Language} & \textbf{Modality}  & \textbf{Participants} & \textbf{Participant Time} & \textbf{Total Time} \\ \midrule
LPPC-fMRI (EN) \cite{RN8}    & English           & fMRI            & 49                     & 1.5 h                     & 73.5 h              \\
LPPC-fMRI (CN)                         & Chinese           & fMRI            & 35                     & 1.5 h                     & 52.5 h              \\
LPPC-fMRI (FR)                         & French            & fMRI            & 28                     & 1.5 h                     & 42.0 h                \\
Chen2026 (EN) \cite{chen2026bilingual}            & English           & fMRI            & 6                      & 2.0 h                       & 12.0 h                \\
Chen2026 (CN)                                  & Chinese           & fMRI            & 6                      & 2.0 h                       & 12.0 h                \\
{SMN4Lang} \cite{RN10}      & Chinese           & fMRI        & 12                     & 5.1 h                       & 60.7 h                \\
NEBULA101 (FR) \cite{rampinini2025nebula101}     & French            & fMRI            & 97                     & 3.6 min                   & 5.8 h              \\
NEBULA101 (EN)                                 & English           & fMRI            & 79                     & 3.6 min                   & 4.7 h              \\ \bottomrule
\end{tabular}
\vbox{
\vspace{0.5em}
\noindent\parbox{\textwidth}{\small The datasets cover Chinese, English and French, with fMRI recordings. Participant numbers refer to the samples analyzed in each language condition. Participant time denotes the duration of narrative stimulus presentation per participant; total time sums these durations across participants.}
}
\label{tab:E1}
\end{table*}

\clearpage

\begin{table*}[]
\centering
\caption{Training and evaluation partitions for the primary datasets.}
\label{tab:E2}
\renewcommand{\arraystretch}{1.2}
\setlength{\tabcolsep}{9pt}

\begin{tabular}{@{}lccccc@{}}
\hline
\textbf{Dataset} & \textbf{Stimuli} & \textbf{Data split} & \textbf{Stimulus index} & \textbf{Words} & \textbf{Time (min)} \\ \hline
\multirow{2}{*}{{LPPC-fMRI (EN)}} & \multirow{2}{*}{9} & Train & Story 1--8 & 13400 & 80 \\
 &  & Test & Story 9 & 1974 & 12 \\
\multirow{2}{*}{{LPPC-fMRI (CN)}} & \multirow{2}{*}{9} & Train & Story 1--8 & 20013 & 85 \\
 &  & Test & Story 9 & 2914 & 13 \\
\multirow{2}{*}{{LPPC-fMRI (FR)}} & \multirow{2}{*}{9} & Train & Story 1--8 & 13581 & 86 \\
 &  & Test & Story 9 & 1810 & 11 \\
\multirow{2}{*}{{Chen2026 (EN)}} & \multirow{2}{*}{10} & Train & Story 1--9 & 20256 & 118 \\
 &  & Test & Story 10 & 1643 & 10 \\
\multirow{2}{*}{{Chen2026 (CN)}} & \multirow{2}{*}{10} & Train & Story 1--9 & 18931 & 117 \\
 &  & Test & Story 10 & 1553 & 8 \\
\multirow{2}{*}{{SMN4Lang}} & \multirow{2}{*}{60} & Train & Story 1--57 & 68647 & 290 \\
 &  & Test & Story 58--60 & 3288 & 13 \\ \hline
\end{tabular}

\vbox{
\vspace{0.5em}
\noindent\parbox{\textwidth}{\small LPPC-fMRI, Chen2026 and SMN4Lang were divided into non-overlapping training and test sets at the story level. Stimulus counts refer to stories. Word counts and durations are summed within each partition. For LPPC-fMRI (CN), each Chinese character is counted as one word; Chen2026 (CN) counts use Chinese word segmentation. For SMN4Lang, counts refer to alphanumeric characters in the released word-level transcripts, excluding punctuation and whitespace.}}

\end{table*}

\clearpage




\begin{figure*}[p]
    \centering
    \includegraphics[width=\textwidth]{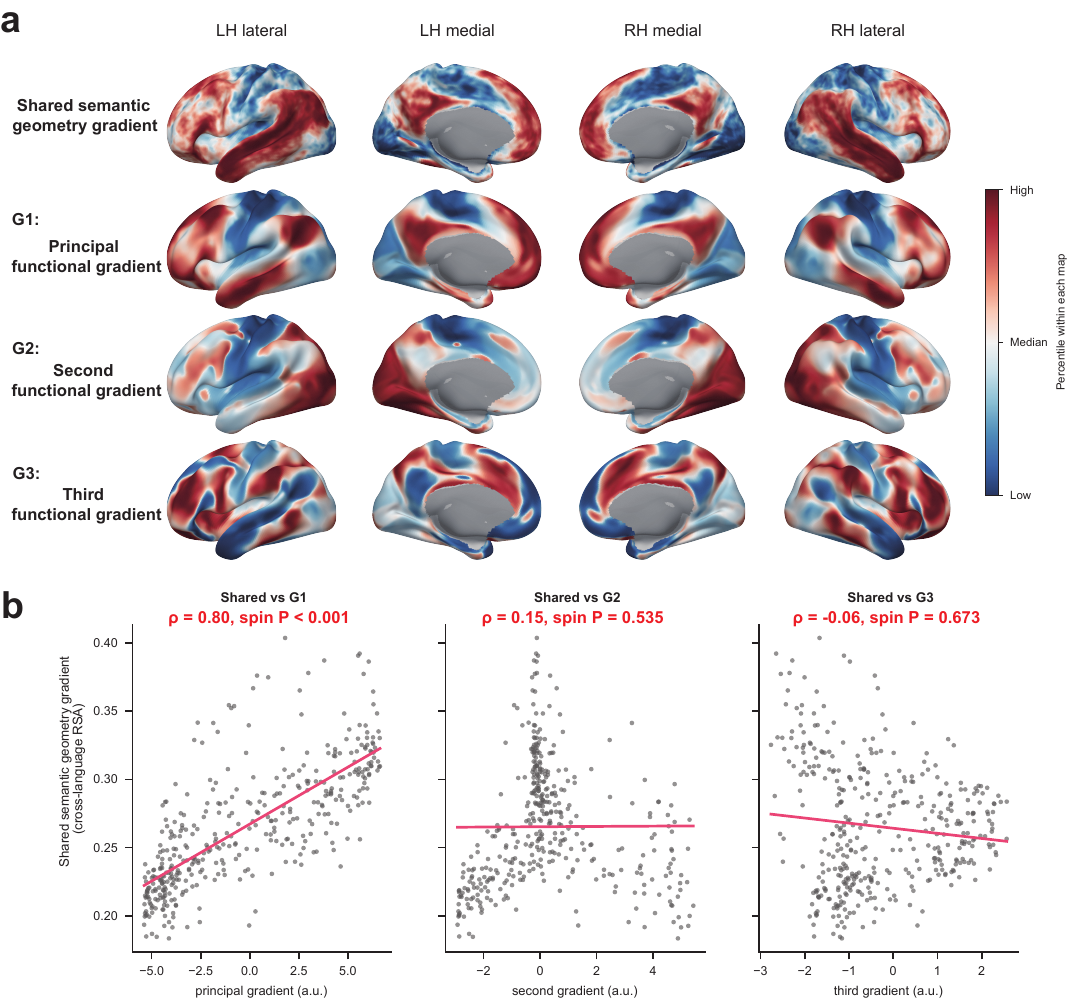}
    \caption{\textbf{Correspondence between shared semantic geometry and cortical functional gradients.} \textbf{a}, Cortical maps of the shared semantic geometry gradient and the first three functional gradients (G1--G3). Shared semantic geometry was defined by cross-language searchlight representational similarity analysis (RSA), averaged across the three language pairs using the LPPC-fMRI training set. Colours indicate percentile ranks within each map. Lateral and medial views of both hemispheres are shown. \textbf{b}, Associations between shared semantic geometry and each functional gradient across Schaefer-400 parcels ($n=400$). Each point denotes a parcel; red lines show least-squares fits. Correlations use raw gradient and RSA values. Spearman correlations and two-sided spin-test $P$ values from 1,000 rotations are indicated. LH, left hemisphere; RH, right hemisphere; a.u., arbitrary units.}
    \label{fig:ex1}
\end{figure*}

\clearpage

\begin{figure*}[p]
    \centering
    \includegraphics[width=\textwidth]{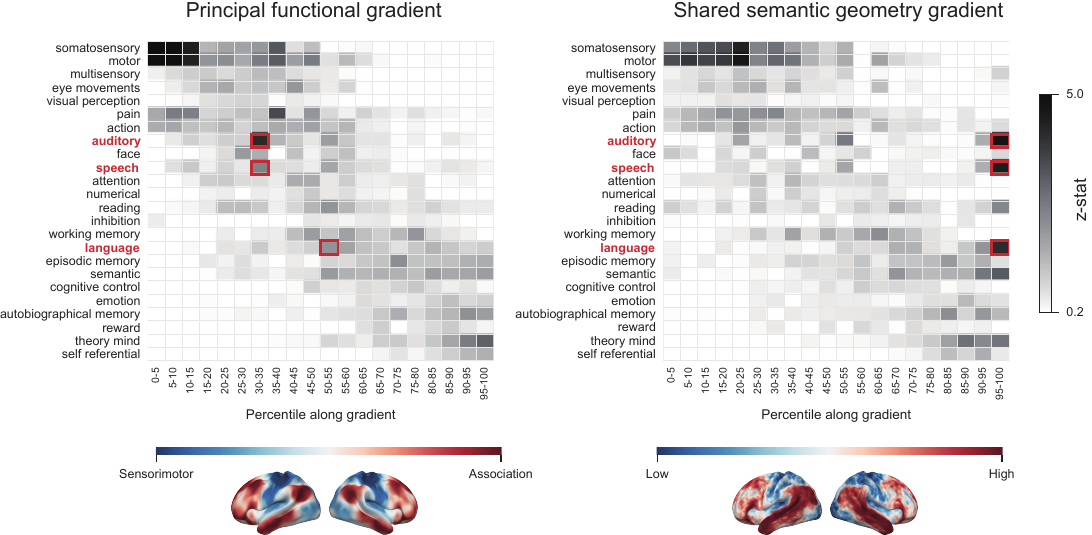}
    \caption{\textbf{Functional associations along the principal functional and shared semantic geometry gradients.} Neurosynth association maps for 24 cognitive terms are summarized along the principal functional gradient (left) and the shared semantic geometry gradient (right). Schaefer-400 parcels were divided into 20 equal-sized percentile bins along each gradient; cells show mean association $z$-statistics, with negative bin means set to zero. Rows are ordered by the weighted mean position of each term along the principal functional gradient and kept identical in both panels. Red boxes mark the peak bins of auditory, speech and language terms. Enrichment of these terms at the apex (top 5\% of parcels) of the shared semantic geometry gradient, relative to the apex of the principal functional gradient, exceeded that of 1,000 surrogate maps accounting for correspondence between the two gradients (auditory, speech and language, all $P=0.001$; one-sided).}
    \label{fig:ex_neurosynth}
\end{figure*}

\clearpage

\begin{figure*}[p]
    \centering
    \includegraphics[width=\textwidth]{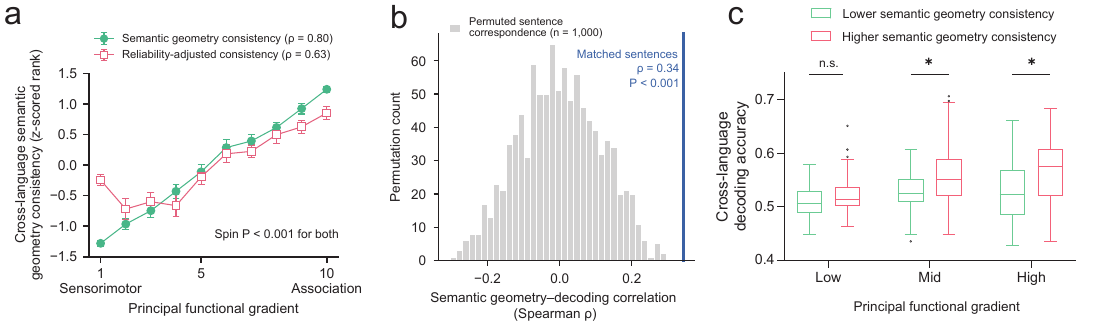}
    \caption{\textbf{Shared semantic geometry and decoding across the cortical functional hierarchy.} \textbf{a}, Semantic geometry consistency across deciles of the principal functional gradient, before and after adjustment for language-specific split-half reliability. Reliability was estimated by correlating sentence representational dissimilarity matrices (RDMs) between participant halves within each language. Original ranks and reliability-adjusted residuals were separately $z$-standardized. Points show means; error bars show standard errors of the mean (s.e.m.) across parcels within each decile. Spearman and reliability-adjusted partial correlations are indicated, with one-sided spin tests using 1,000 rotations. \textbf{b}, Null distribution of the association between semantic geometry consistency and single-region cross-language decoding after 1,000 within-story permutations of cross-language sentence correspondence (grey histogram). The blue line marks the association for matched sentences. \textbf{c}, Single-region zero-shot cross-language decoding in parcels with lower or higher semantic geometry consistency within each third of the principal functional gradient (Low, Mid and High). Parcels within each third were divided at the median semantic geometry consistency. Boxes summarize parcel-wise accuracies; centre lines indicate medians and boxes indicate interquartile ranges. Within each third of the principal functional gradient, higher and lower semantic geometry consistency groups were compared using one-sided spin tests (1,000 rotations), with Holm correction across the three comparisons; asterisks indicate adjusted $P$ values: *$P<0.05$. Analyses used Schaefer-400 parcels in LPPC-fMRI. Semantic geometry consistency was averaged across three language pairs using the training set. Decoding in \textbf{b,c} used two-alternative forced-choice identification with one parcel retained at a time, averaged across six language directions.}
    \label{fig:ex3_functional_gradient_controls}
\end{figure*}

\clearpage

\begin{figure*}[p]
    \centering
    \includegraphics[width=\textwidth]{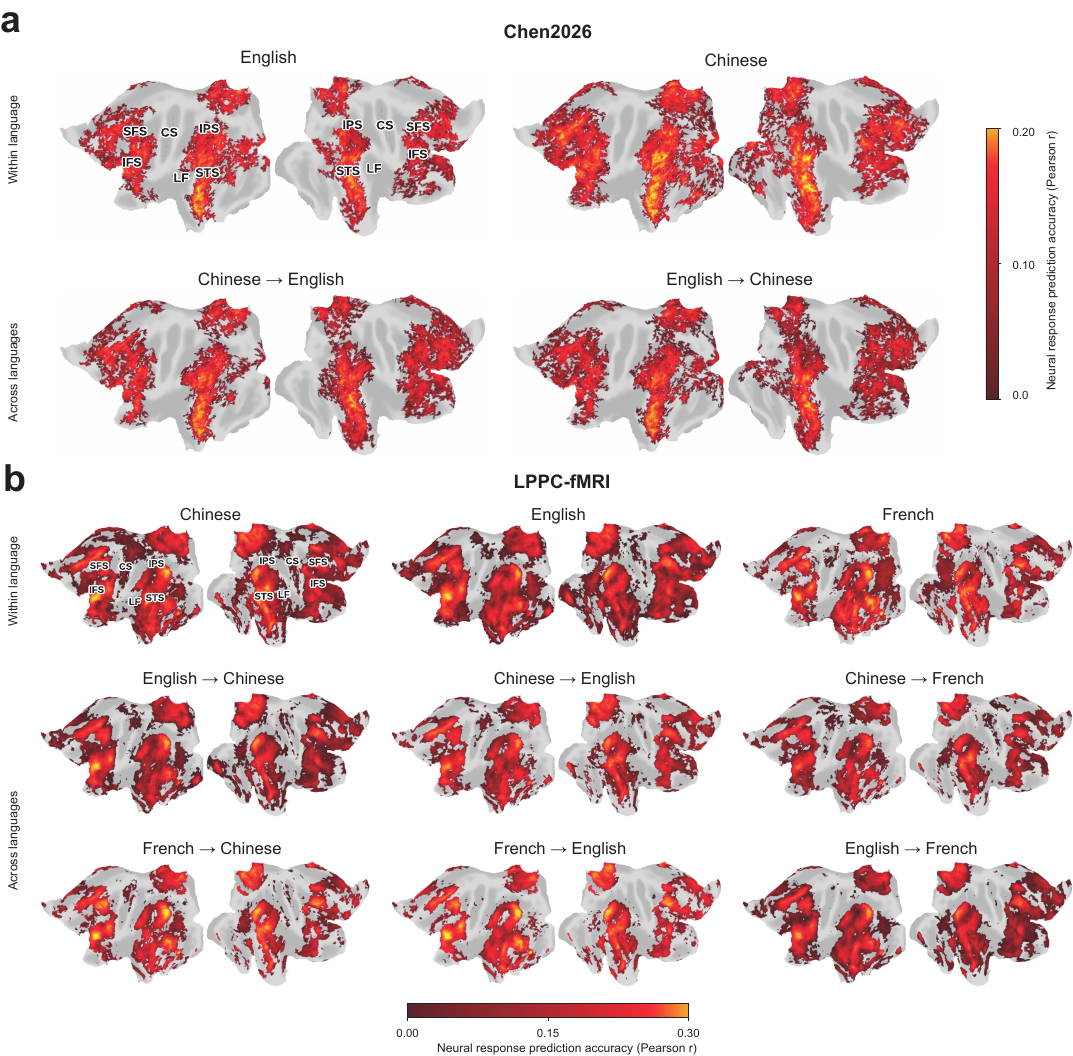}
    \caption{\textbf{Within-language and cross-language neural response prediction.} Models were fitted separately for each source participant using leave-one-story-out cross-validation. \textbf{a}, Neural response prediction within English and Chinese (top row) and across both language directions (bottom row) in Chen2026. Models were fitted and evaluated within each bilingual participant. \textbf{b}, Neural response prediction within Chinese, English and French (top row) and across all six language directions (middle and bottom rows) in LPPC-fMRI. Predictions were compared with target-language group-mean responses; within-language evaluation excluded the source participant from the group mean. Arrows indicate training to test language. Colours on flattened cortical surfaces show Pearson correlations between predicted and observed responses, with separate scales for the two datasets. Only correlations surviving one-sided sign-flip tests with family-wise error correction are coloured ($P<0.05$; vertex-level max-$T$ in LPPC-fMRI, cluster-mass correction in Chen2026). SFS, superior frontal sulcus; IFS, inferior frontal sulcus; CS, central sulcus; IPS, intraparietal sulcus; STS, superior temporal sulcus; LF, lateral fissure.}
    \label{fig:ex2}
\end{figure*}

\clearpage

\begin{figure*}[p]
    \centering
    \includegraphics[width=\textwidth]{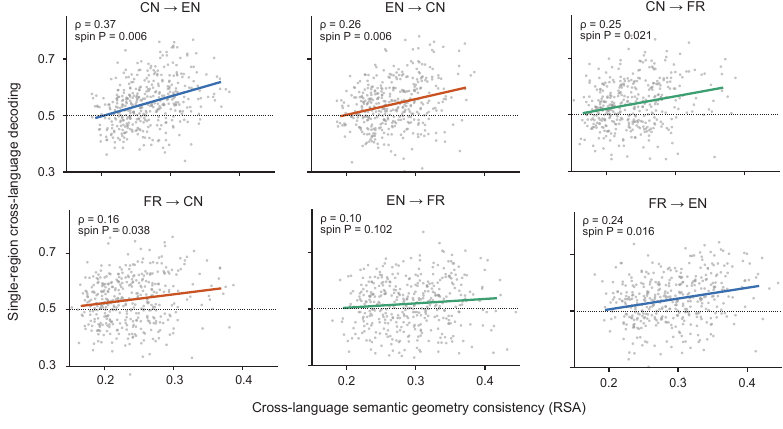}
    \caption{\textbf{Regional semantic geometry consistency and cross-language decoding.} Associations between cross-language RSA and single-region decoding accuracy in LPPC-fMRI, shown for each of the six directions of cross-language decoding. Arrows indicate training to test language. Each point denotes a Schaefer-400 parcel ($n=400$). Decoding used two-alternative forced-choice identification, retaining only the indicated parcel's standardized cortical input. Solid lines show least-squares fits; dotted lines indicate chance accuracy (0.5). Spearman correlations and one-sided spin-test $P$ values (1,000 rotations), Holm-corrected across the six directions, are indicated. RSA, representational similarity analysis; CN, Chinese; EN, English; FR, French.}
    \label{fig:ex_lppc_geometry_decoding}
\end{figure*}

\clearpage

\end{document}